\documentstyle[12pt]{article}
\begin{document}
\def\theequation{\arabic{section}.\arabic{equation}}
\begin{titlepage}
\title{Non--minimal coupling of the scalar field and inflation}
\author{Valerio Faraoni \\ \\
{\small \it Department of Physics and Astronomy, University 
of Victoria} \\
{\small \it P.O.
 Box 3055, Victoria, B.C. V8W 3P6 (Canada)}}
\date{}
\maketitle  % \thispagestyle{empty}  
\vspace*{1truecm}
\begin{abstract} 
We study the prescriptions for the coupling constant of a scalar field to the
Ricci curvature of spacetime in specific gravity and scalar field theories. 
The results are applied to the most popular
inflationary scenarios of the universe; their theoretical consistency and
certain observational constraints are discussed.
\end{abstract}
\vspace*{1truecm} 
\begin{center}  
To appear in {\em Physical Review D}
\end{center}     
\end{titlepage}   \clearpage

\section{Introduction}

The concept of inflation has dominated the cosmology of the early universe for
the last fifteen years. Despite the success of the inflationary paradigm in
resolving the problems of the standard big--bang model and in providing a
mechanism for the formation of structures in the universe, there is no
universally accepted model for inflation: rather, many different 
inflationary scenarios have been proposed. Moreover, it has not 
been possible to
unambigously identify the inflaton with any known field from a particle 
physics theory. A comparison of the inflationary models 
with observations has been made possible in recent years by the discovery 
of anisotropies
in the cosmic microwave background \cite{Smootetal}. A
difficulty that is often encountered in comparing theory and observations 
is that a specific inflationary scenario
typically contains several free parameters, and a {\em ad hoc} choice of their
values may render the scenario viable, sometimes at the price of 
fine--tuning the
parameters or the initial conditions of the model (see e.g.
Refs.~\cite{FutamaseMaeda89,MakinoSasaki91,ALO90}). In the 
present paper, we study
the possible prescriptions for one of the parameters appearing in many
inflationary scenarios, namely the coupling constant $\xi$ of the inflaton 
with
the Ricci curvature of spacetime. To fix the ideas, let us consider the
Lagrangian density for Einstein gravity and a non--minimally coupled scalar
field as the only form of matter:\setcounter{equation}{0}
\begin{equation}  \label{Lagrangiandensity}
{\cal L}=\left[ \frac{R}{16\pi G}-\frac{1}{2} \nabla^{\mu}\phi  
\nabla_{\mu} \phi -V( \phi)-\frac{m^2}{2} \phi^2
- \frac{\xi}{2}R\phi^2 \right] \sqrt{-g}\; ,
\end{equation}
where $R$ denotes the Ricci curvature of spacetime, $g$ is the determinant of
the metric $g_{\mu\nu}$, $\nabla_{\mu}$ is the covariant derivative operator,
$m$ and $V( \phi)$ are, respectively, the mass and the potential of 
the scalar field $\phi$. $\phi$ obeys the
Klein--Gordon equation
\begin{equation}   \label{KG}
\Box \phi-\xi R \phi-m^2\phi -\frac{dV}{d\phi}=0 \; .
\end{equation}
The term $-\xi R \phi^2 /2$ in the Lagrangian density 
(\ref{Lagrangiandensity}) describes the non--minimal coupling of the field
$\phi$ to the
curvature \cite{footnote1}. 
It is well--known \cite{Abbott81,FutamaseMaeda89,ALO90} that the viability of
inflationary models is deeply affected by the value of the
parameter $\xi$. Although a
popular choice is setting $\xi=0$ (minimal coupling) in order to simplify the
calculations, this prescription for $\xi$ is often unacceptable. 
In quantum field theory in curved spacetimes it is argued 
that a non--minimal coupling is to be expected when the
spacetime curvature is large. Non--minimal couplings are generated by quantum
corrections even if they are not present in the classical action
\cite{Linde82}. The coupling is actually required if the scalar field theory 
is to be renormalizable in a classical gravitational background
\cite{FreedmanMuzinichWeinberg74,FreedmanWeinberg74}. When the problem of the
correct value of $\xi$ is not
ignored, the prevailing point of view in the 
literature on inflation is that the
coupling constant $\xi$ is a free parameter, and that the values of $\xi$
that are acceptable are those that, {\em a posteriori}, make a specific
inflationary scenario viable. In this paper, we show that this 
point of view is
unacceptable in many cases, and that often there exist definite prescriptions
for the coupling constant. The value of $\xi$ depends on the nature of the
inflaton $\phi$ and on the theory of gravity under consideration. With the 
value of $\xi$ known {\em a priori}, specific scenarios 
are analyzed and their 
theoretical consistency is discussed, before comparing their predictions 
with the available observations. 

The plan of the paper is as follows: in Sec.~2 we illustrate the various
prescriptions for the value of $\xi$ in different theories, and 
we study their
applicability to inflation. Emphasis is given 
to metric theories of gravity, in
particular general relativity and theories formulated in 
the Einstein conformal
frame. In Sec.~3 we examine the consequences of these prescriptions for the
most popular inflationary scenarios proposed so far. Section~4 contains 
considerations on the effects of non--minimal coupling in power--law inflation
and observational constraints on a specific model. In Sec.~5 we provide
further constraints on chaotic and new inflation. Section~6 contains the
conclusions.

\section{Prescriptions for the coupling constant $\xi$}

The coupling constant $\xi$ is often regarded as a free parameter 
in inflationary scenarios. This view arises from the fact that there is no
universal prescription for the value of $\xi$. Indeed, some
precriptions for $\xi$ do exist in specific theories, although they are 
not widely known, and
they depend on the nature of the scalar field $\phi$ and on the 
theory of gravity. 
In this section, we will review the prescriptions for the coupling constant,
before applying them to cosmology in Sec.~3.

\subsection{Quantum theories of the scalar field $\phi$}

The available prescriptions for the coupling constant 
$\xi$ differ depending on
whether the scalar is a fundamental field, or is associated with a composite
particle. In Ref.~\cite{VoloshinDolgov} it was argued that, if $\phi$ is a
Goldstone boson in a theory with a spontaneously broken global symmetry, 
then $\xi=0$. It has been pointed out that if the scalar field
$\phi$ is associated to a composite particle, the value of $\xi$ should be
fixed by the known dynamics of its constituents \cite{HillSalopek92}. In
particular, in Ref.~\cite{HillSalopek92}, the Nambu--Jona--Lasinio model was
analyzed and, in the large $N$ approximation, the value $\xi=1/6$ 
was found for this specific model. Reuter \cite{Reuter94} considered the 
$O(N)$--symmetric model with a quartic self interaction, in which the
constituents of the $\phi$ boson are scalars themselves. The resulting $\xi$
depends on the coupling constants of the elementary scalars \cite{Reuter94}.
Other arguments restrict the range of allowed values of $\xi$; Hosotani
\cite{Hosotani85} examined the 
back reaction of gravity on the stability of the
scalar field $\phi$ assuming the Lagrangian of Einstein gravity with a general
coupling $\xi R \phi^2/2$ and a potential\setcounter{equation}{0}
\begin{equation}
 V( \phi)=V_0+\frac{m^2}{2}\, \phi^2+\frac{\eta}{3!}\,\phi^3
+\frac{\lambda}{4!}\,\phi^4 \; .
\end{equation}
He found that, for cubic self--interactions, $\xi=0$ is the only value allowed.
For Higgs scalar fields in the standard model \cite{footnote2}, it 
must be $\xi \leq 0$ or $\xi \geq 1/6$. However, the results of
Ref.~\cite{Hosotani85} are based on the use of canonical gravity and the
conclusions may change if an alternative theory is adopted for the background
gravity. 

To our knowledge, no other prescriptions for the coupling constant $\xi$ are 
available from quantum
theories of the field $\phi$. It is likely that every theory
which provides a candidate for the inflaton will provide a specific value, or
range of values, for $\xi$. To make things worse, in a quantum theory $\xi$ is
subject to renormalization, like masses or other coupling constants
\cite{Reuter94,ParkerToms85}. It appears, therefore, that the prospects for an
unambigous determination of $\xi$ are not promising. However, this would be a
pessimistic conclusion, because inflation is essentially a {\em classical},
low energy, phenomenon. It has been 
argued that ``the tensor contribution to the
cosmic microwave background quadrupole implies that the vacuum energy that
drives inflation is not a quantum--gravitational phenomenon''
\cite{Turnerreview}. To be more specific, the potential energy density of the
scalar field 50 e--folds before the end of inflation is subject to the
constraint $V_{50} \leq 6 \cdot 10^{-11}m_{pl}^4$ \cite{Turnerreview}. 
Hence, gravity is classical during inflation.
In many scenarios, the inflaton $\phi$ is a {\em gravitational} field,
(e.g. the field of Brans--Dicke theory), and hence it is classical. 
What if the
inflaton is non--gravitational in origin~? The problem whether a classical
treatment of the inflaton is appropriated has been studied in a number of
papers (\cite{cinflaton,Lyth85,Sasaki86} and references therein). Under 
certain conditions, the 
distribution of the
field is peaked around classical trajectories and the evolution of the scalar
field can be considered as classical. This justifies the use of {\em classical}
equations to describe inflation, and it is not inconsistent with the fact that
quantum fluctuations of $\phi$ around its classical value provide seeds for
density perturbations \cite{Lyth85,Sasaki86}. Therefore, the problem of the
determination of the correct value of the coupling constant $\xi$ may be 
restricted to the consideration of classical theories of gravity {\em and} 
of the inflaton $\phi$. 

\subsection{Classical theories of $\phi$ and metric theories of gravity}

According to the previous discussion, we will assume 
that gravity is described
by a classical theory based on a spacetime manifold, and that the inflaton 
field $\phi$ is classical. There exists a prescription for 
the coupling constant
$\xi$ of a scalar field with the Ricci curvature, and for the coupling 
constants 
with other curvature scalars which can, in principle, be considered. The
generalization of the flat space Klein--Gordon equation to a curved spacetime
includes couplings with the Ricci curvature, as well as couplings with the
other scalars constructed from the curvature tensor:
\begin{equation}   \label{generalKG}
\Box \phi -m^2 \phi - \left( \xi R +\alpha_1 R^2+\alpha_2
R^{\alpha\beta}R_{\alpha\beta}+\alpha_3 R^{\alpha\beta\gamma\delta}
R_{\alpha\beta\gamma\delta}+... \right) \phi-\frac{dV}{d\phi}=0 \; .
\end{equation}
In Ref.~\cite{SFFS} it was proved that, under the assumptions
\begin{itemize}
\item {\em i}) the scalar field $\phi$ satisfies Eq.~(\ref{generalKG});
\item {\em ii}) the field $\phi$ satisfies the Einstein equivalence 
principle (hereafter EEP--see Ref.~\cite{Will} for a formulation), i.e. 
the propagation of $\phi$ resembles locally the propagation in flat
space;
\end{itemize}
the coupling constants are forced to assume the values
\begin{equation}        \label{GRprescription}
\xi=1/6 \;\;\;\; , \;\;\;\; \alpha_1=\alpha_2=\alpha_3=...=0 \; .
\end{equation}
This result arises from the study of wave propagation and 
tails of radiation in
a curved spacetime, and was derived by requiring that the structure 
of tails of
radiation become closer and closer to that occurring in flat spacetime when
the curved manifold is progressively approximated 
by its tangent space (i.e. by
imposing the EEP on the field $\phi$). Although the requirement of 
Eq.~(\ref{GRprescription}) reproduces the usual case of conformal coupling in
four spacetime dimensions, the derivation of this result is
completely independent of conformal transformations, the conformal 
structure of
spacetime, the particular spacetime metric and the field equations for the
metric tensor in the particular theory \cite{SFFS}. The conclusions of
Ref.~\cite{SFFS} were confirmed in Ref.~\cite{GribPoberii}.

If the assumption {\em i)} is satisfied but {\em ii)} is not there is, in
principle, the disturbing possibility  that massive scalar particles 
propagate on the light cone in a space in which the Ricci curvature is
different from zero \cite{SFFS}.

A question arises naturally: can we impose the EEP on the scalar field
(inflaton) $\phi$ in a particular theory of gravity (inflationary scenario)~?
The answer depends on the gravitational theory under consideration. If 
the nature of the field $\phi$ is gravitational (e.g. the scalar field of
Brans--Dicke theory), the statement that its physics resembles locally the
physics in flat spacetime goes beyond the EEP, which regards only {\em
non--gravitational} physics \cite{Will}. Metric 
theories of gravity \cite{Will}
(including general relativity--hereafter GR) satisfy the EEP. Therefore, if 
the correct  theory of gravity during inflation is GR, or any metric theory 
in which
the inflaton field $\phi$ is non--gravitational, then the EEP holds and the
coupling constant assumes the value $\xi=1/6$. GR is widely 
used in the construction of
inflationary scenarios, but it is not the only theory used for this purpose.
Almost all the existing scenarios of inflation employ a metric theory of
gravity: however, the inflaton field $\phi$ can have 
gravitational origin, like
in scalar--tensor theories (of which Brans--Dicke theory is the simplest 
example).
The prescription $\xi=1/6$ clearly does not apply to the latter case 
for the above mentioned reason \cite{footnote3}, 
and also for a second reason: the field $\phi$ in these
theories satisfies an equation more complicated than Eq.~(\ref{generalKG})
\cite{Will}. 
However, if the field $\phi$ satisfies an equation of the
kind~(\ref{generalKG}) and is massive ($m\neq 0$ is assumed in 
many inflationary
scenarios), any value of $\xi$ different from $1/6$ leaves the possibility 
that the massive $\phi$ propagates along the light cones when $R\neq 0$
\cite{SFFS}. This argument supports the choice $\xi=1/6$ for {\em any} massive 
field satisfying Eq.~(\ref{generalKG}). However, in the following, 
we will regard the prescription~(\ref{GRprescription}) as 
valid only for GR and all the metric theories of gravity in which the inflaton
field $\phi$ is non--gravitational.

\subsection{Theories formulated in the Einstein frame}

A wide class of theories of gravity can be grouped into this category. 
They have
the common feature that the final formulation of the theory is made in the
``Einstein frame'' conformally related to the ``Jordan frame'', in which
the theory was formulated at the start (for definitions and terminology,
we refer the reader to \cite{MagnanoSokolowski} and references therein).
This class of theories includes Kaluza--Klein, $R^2$, 
supergravity and string--inspired theories, and many 
generalized scalar--tensor theories. The conformal
transformation to the Einstein frame has also been used as a mathematical 
technique to transform a non--minimally coupled scalar field to the
(computationally much easier) case of a minimally coupled field. In the
literature, there is plenty of ambiguity on which conformal frame should be
regarded as physical. For some theories, it has been proved that 
the ``original''
formulation in the Jordan frame is physically unacceptable 
because the kinetic
energy of the scalar field is negative--definite, and a unique conformal 
transformation
to the Einstein frame is singled out. In these cases, the formulation 
in the Einstein
frame is the only acceptable possibility. The necessity (and uniqueness) 
of the conformal transformation has been established for 
Brans--Dicke \cite{footnote4} 
and Kaluza--Klein theories \cite{ConfTrans,Cho92}, and 
has been generalized to
a wider class of
theories \cite{MagnanoSokolowski}. The theory in the Einstein frame is,
in general, very different from the Jordan frame formulation. The conformal
transformation to the Einstein frame (and the associated redefinition of the
scalar field~--~see \cite{Cho92,MagnanoSokolowski}) has 
the consequence that the
``new'' scalar field in the Einstein frame is minimally coupled to the
curvature, $\xi=0$, irrespective of the value of the coupling constant in the
Jordan frame. This prediction applies to all theories
formulated in the Einstein frame, which have been used 
extensively to construct
inflationary cosmologies.

We remark that, according to 
Ref.~\cite{SFFS}, the minimally coupled scalar field of a theory formulated
in the Einstein frame violates the EEP. Therefore, strictly 
speaking, the theory
is not Einstein gravity, in which the EEP (and also the strong equivalence
principle) are satisfied. This fact conflicts with the current use of the 
term ``Einstein gravity'' in many papers. The
violation of the EEP in a theory formulated in the Einstein frame is not
surprising, since also the weak equivalence principle is violated in these
theories. In fact, if a form of matter (let us say a
field $\psi$, to fix the ideas) other than the inflaton is included in
the Jordan frame Lagrangian, then the stress--energy tensor of $\psi$ 
in the Einstein frame is non--minimally coupled to the inflaton. This causes
the presence of a fifth force violating the equivalence principle
\cite{Cho92} and a time dependence of the coupling constants 
of physics, which is
actually regarded as an important low--energy manifestation of 
string (and other)
theories. When the weak equivalence principle is violated by the conformally
transformed field $\psi$, the violation of the EEP (a stronger version 
of the equivalence principle) by the inflaton does not appear to be 
surprising. It is to be
noted that if the scalar field decays and disappears from the universe during
the radiation era, or at an early time during the matter--dominated 
era \cite{DamourNordvedt} (or even earlier \cite{GBW}), the violation of
the equivalence principle leaves no trace in present day experiments performed
in the Solar System.
%%%%%%%%%%%%%%%%%%%%%%%%%%%%%%%%%%%%%%%%%%%%%%%%%%%%%%%%%%%%%%%%%%%%%%%%%%%%
\section{Consequences of the prescriptions of $\xi$ for inflation}

In this section, we apply the predictions of Sec.~2 to cosmology. We 
examine the inflationary scenarios most studied in the
literature and we answer the two following questions: 1) is any of the
prescriptions examined in Sec.~2 for the value of $\xi$ applicable ?
and 2) if the answer to question 1) is affirmative, what are the consequences 
for the specific inflationary scenario~?

It is well--known that the viability of a particular inflationary model can
depend strongly on the value of the coupling parameter $\xi$. The following
arguments have been used to argue against or in favour of specific scenarios:
\begin{itemize}
\item the existence of inflationary solutions;
\item the amount of inflation necessary to solve the problems of the standard
big--bang model;
\item the fine--tuning of initial conditions for the inflaton.
\end{itemize}
These conditions regard the unperturbed model of the universe. A fourth
argument to be taken into account is the evolution of density perturbations
generated during inflation.

Many results on the viability of inflationary scenarios with a non--minimally
coupled scalar field are already available in the literature. In these
papers, the choice of the value of $\xi$ 
was motivated {\em a posteriori} by the
viability of the inflationary scenario, according to the prevailing point of
view that sees $\xi$ as a free parameter. Our point of
view is radically different from previous works: While $\xi$ was a
free parameter for the previous authors, we have the prescription $\xi=1/6$
for the metric theories of gravity in 
which the inflaton is non--gravitational.
We review the results available in the literature from our new point of view. 
The scenarios analyzed in the following are the most well--studied, but do not
constitute a complete list of the models proposed in the literature.

\subsection{New inflation}

The new inflationary scenario \cite{LindePLB82,AlbrechtSteinhardt82} 
currently 
is not regarded as a successful one because of the extreme fine--tuning of
parameters in the effective potential required to reproduce the observable
universe \cite{BardeenSteinhardtTurner83,GuthPi82,Hawking82}. However, the 
study of new inflation provides insight in the way non--minimal 
coupling affects a slow--roll
inflationary scenario. The background gravity for new inflation is 
assumed to be GR and the (unperturbed) inflaton field $\phi$ 
is treated as classical,
and is non--gravitational. The prescription $\xi=1/6$ applies. Abbott
\cite{Abbott81} considered this scenario with the Coleman--Weinberg 
potential\setcounter{equation}{0}
\begin{equation}    \label{ColemanWeinberg}
V(\phi)=B\phi^4 \left[ \ln \left( \frac{\phi^2}{\sigma^2} \right)
-\frac{1}{2} \right] +\frac{B\sigma^4}{2} \; ,
\end{equation}
where $B$ is constant and $\sigma=10^{15}$~GeV, and realized that, if
$\xi>0$, the term $\xi R \phi^2/2$ in the 
Lagrangian density acts like an extra
term in the scalar field potential, and creates a 
barrier that prevents the GUT
phase transition from being completed. During the slow--roll 
of the inflaton on
the flat section of the potential, the universe behaves very much like de
Sitter space ($R=$constant), and the term $\xi R \phi^2/2$ 
behaves like a mass
term for the scalar field \cite{footnote5}, destroying the 
flatness of the potential. This
happens, in particular, for $\xi=1/6$. It is to be concluded 
that this version
of the new inflationary scenario is not theoretically 
consistent, regardless of
the fine--tuning problems. 

Flat potentials different from 
(\ref{ColemanWeinberg}) can also achieve new inflation. For example, the
potentials
\begin{equation}
V( \phi)=V_0-\alpha \phi^2-\beta \phi^3+\lambda \phi^4 \; ,
\end{equation}
\begin{equation} 
W( \phi)=W_0-\alpha \phi^4+\beta \phi^6         \; ,
\end{equation}
(where $V_0$, $W_0$, $\alpha$, $\beta$ and $\lambda$ are constants) and 
have been employed
\cite{SteinhardtTurner84,Turnerreview}. The effect of non--minimal coupling
appears to be the same in this potential as in the Coleman--Weinberg 
potential. In general, {\em the argument of 
Ref.~\cite{Abbott81} applies to all scenarios with a slow--rollover 
inflationary potential: the flatness of the potential is destroyed by the 
non--minimal coupling of the inflaton.} What about non flat 
potentials (used, e.g., in chaotic
or power law inflation)~? In principle there is the possibility that the term 
$\xi R \phi^2/2$ in the Lagrangian density balances a suitable 
potential $V( \phi)$ in
such a way that a section of the resulting ``effective potential'' is almost
flat, thus giving again slow--roll of the inflaton field. The energy density
and pressure of a non--minimally coupled scalar field are given by
\begin{equation}   \label{density}
\rho=\left( 1-8\pi G \xi \phi^2 \right)^{-1} \left[ \frac{(
\dot{\phi})^2}{2}+V( \phi)+6\xi H \phi \dot{\phi}\right] \; ,
\end{equation}
\begin{equation}                  \label{pressure}
P=\left( 1-8\pi G \xi \phi^2 \right)^{-1} \left[ \left( \frac{1}{2}-2\xi 
\right) 
\dot{\phi}^2-V( \phi)-2\xi \phi \ddot{\phi}-4\xi H \phi \dot{\phi} \right] \;.
\end{equation}
It is possible to consider a suitable potential 
such that the equation of state
approaches $P=-\rho$ and thus achieve inflation in the presence of 
a non--minimal coupling. A concrete example was given in
Ref.~\cite{BayinCooperstockFaraoni} in the context of GR with 
conformal coupling 
($\xi=1/6$), by assuming the equation of state $P=( \gamma-1) \rho$ 
and deriving
numerically (for small values of the constant $\gamma$) the 
necessary potential. This potential is very different 
from the corresponding potential derived
analytically for $\xi=0$ for the same values of $\gamma$ in
Ref.~\cite{StarkovichCooperstock}. Unfortunately, when the ``effective 
potential'' 
$V+\xi R \phi^2/2$ has a flat section on which the inflaton 
rolls slowly, the
complication of the Friedmann and the 
Klein--Gordon equations prevents us from
developing an elegant slow--roll formalism in terms of slow--roll parameters
like the one available for a minimally coupled scalar field 
\cite{SteinhardtTurner84}. The introduction of an 
effective potential mimicking
the effects of non--minimal coupling does not appear to 
be possible, as will be
shown in Sec.~4.

\subsection{Power--law inflation}

For a minimally coupled scalar field, power--law inflation
\cite{AbbottWise84,LucchinMatarrese85,Barrow87,Liddle89} arises from an
exponential potential 
\begin{equation} \label{PLIpotential}
V( \phi)= V_0 \exp \left( \pm \sqrt{ \frac{16\pi}{p}} \frac{\phi}{m_{pl}}
\right) \; ,
\end{equation}
where $p>1$ is constant and the scale factor has the time dependence
\begin{equation} \label{PLI}
a(t) =a_0\, t^p   \; .
\end{equation}
It was
recognized in Ref.~\cite{LucchinMatarrese85} that the potential 
(\ref{PLIpotential}) 
is motivated in the context of Kaluza--Klein cosmologies. Actually, 
exponential
potentials arise in string theories, 
supergravity and, in general, in any
theory which is obtained by means of a conformal 
transformation to the Einstein
frame. As discussed in Sec.~2.3, the prescription $\xi=0$ is the only
possibility in this case. The expression 
``power--law inflation'' generically denotes a scenario in which the 
scale factor has the time--dependence (\ref{PLI}), rather than 
a realization of inflation in a
particular theory of particle physics. Most of the times, the particular 
theory in which power--law inflation is considered is not specified in the
literature. We conclude that the power--law inflationary scenarios 
based on a theory formulated in the Einstein frame are theoretically 
consistent only if $\xi=0$. Examples are given by the class of 
models \cite{footnote6} (representative of 
Kaluza--Klein cosmologies and other theories) of Ref.~\cite{Cho90ChoYoon}, 
and by extended
inflation reformulated in the Einstein frame \cite{KolbSalopekTurner90}. 

\subsection{$R^2$ inflation}

Higher derivative theories of gravity 
 have the peculiar feature that inflation is
generated by the $R^2$ term in the Lagrangian density for gravity, and a 
scalar
field is not needed \cite{Starobinski80,Starobinski84}, 
thus bypassing the problem of the value of $\xi$. However,
a scalar field is sometimes included in the scenario to ``help'' inflation 
(see e.g. Ref.~\cite{Maeda89}).
Since gravity is not Einstein gravity, a prescription for the coupling
constant $\xi$ is not available. To give an idea, we consider the 
proposed form of the Lagrangian density:
\begin{equation}
L=\frac{m_{pl}^2}{16\pi} \left( R +\frac{R^2}{6M^2} \right)
+L_{non-gravitational} \; ;
\end{equation}
the justification for this Lagrangian density comes from 
supergravity \cite{CardosoOvrut93}. There is no point in imposing the EEP in
the context of supergravity: in fact it is known that already the weak
equivalence principle is violated at least in $N=2$ and $N=8$ supergravity 
\cite{Scherk} even
in the low energy, weak field limit, with consequences testable by current
experiments (which are actually used to constrain these theories
\cite{BellucciFaraoni}). In any case, it appears that both isotropic
and anisotropic cosmologies have inflationary solutions as attractors,
irrespective of the value of $\xi$ \cite{MaedaSteinFutamase89}.

By means of a conformal transformation to the Einstein frame, $R^2$ inflation
can be recast as ``standard'' gravity with a minimally coupled scalar field
\cite{LiddleLythreview}. This version of the theory is theoretically
consistent. 

\subsection{Extended and hyperextended inflation}

Extended inflation in its original formulation \cite{LaSteinhardt89} made use
of Brans--Dicke theory; the
original scenario was soon abandoned due to the ``big bubble problem''.
Extended inflation can be recast as power--law inflation after a conformal
transformation to the Einstein frame, with $p$ in Eq.~(\ref{PLI}) given by 
$p=\frac{\omega}{2}+\frac{3}{4}$
(where $\omega$ is the Brans--Dicke parameter) \cite{KolbSalopekTurner90}. In
this formulation, the scenario is theoretically consistent.

A version of extended inflation in which the inflaton $\chi$ is different from
the Brans--Dicke field $\phi$ and is coupled non--minimally to the spacetime 
curvature
($\xi_{\chi}\neq 0$) has been proposed \cite{LaycockLiddle94}. The 
field $\chi$
is non--gravitational and Brans--Dicke theory is a metric theory of gravity,
hence the prescription $\xi_{\chi} =1/6 $ applies. However, there 
are two other
parameters ($\chi_0$, $\omega$), which make difficult to draw 
conclusions on the
viability of this scenario, and a conformal transformation to the Einstein
frame may be necessary \cite{MagnanoSokolowski}.

Hyperextended inflation
\cite{SteinhardtAccetta90,LiddleWands92,CrittendenSteinhardt92,GarciaQuiros90}
is based on scalar--tensor theories that generalize Brans--Dicke theory. The
inflaton is a gravitational scalar field which is not subject to
the prescription $\xi=1/6$. $\phi$ is directly coupled to the Ricci
curvature via a term $Rf( \phi)$, where $f( \phi)$ is an arbitrary function of
$\phi$, and the equation satisfied by $\phi$ is different from Eq.~(\ref{KG}).

\subsection{Induced gravity inflation}

Induced gravity inflation 
\cite{AccettaZollerTurner85} is also based on a scalar--tensor theory and the
inflaton has gravitational origin. A
non--minimal coupling $\xi \neq 0$ has been used \cite{Spokoiny84} 
in conjunction with the
Coleman--Weinberg potential (\ref{ColemanWeinberg}). Chaotic
inflation has been achieved in the context of induced gravity
\cite{CervantesDehnen95}. No prescription for $\xi$ is available in these
cases. Induced gravity inflation has also been reformulated in the Einstein
frame \cite{SalopekBondBardeen89}; this scenario with $\xi=0$ 
is theoretically consistent, as explained in Sec.~2.3.

\subsection{Chaotic inflation}

Several results on chaotic inflation with a non--minimal
coupling are available in the literature. The chaotic 
inflationary scenario originally introduced by Linde
\cite{Linde83} employs GR and the Einstein equations are generally used in
papers on the subject (e.g. \cite{Linde2Mezhlumian95}). 

Futamase and Maeda \cite{FutamaseMaeda89} considered this scenario with a
massive or massless scalar field, with the potential 
\begin{equation} \label{quarticpotential}
V( \phi)=\lambda \phi^4
\end{equation}
(the same potential
originally introduced by Linde \cite{Linde83}) and a non--minimal coupling 
of the inflaton, $\xi\neq 0$. They found that if $\xi
\geq 10^{-3}$, chaotic inflation requires fine--tuning in the initial
conditions for the scalar field. They concluded that ``the
chaotic scenario in the non--minimally coupled model does not work unless the
coupling constant $\xi$ is negative or sufficiently small ($\xi \leq
10^{-3}$)~... Thus, in order to know 
whether the chaotic inflationary scenario
does really work, one has to investigate first whether the inflaton couples
minimally or nonminimally with the spacetime curvature. If it turns out that
the inflaton couples nonminimally classically or possibly through quantum
corrections, one has to investigate how strong the coupling is'' 
\cite{FutamaseMaeda89}. 
It is clear from our discussion of Sec.~2.2 that the scenario 
considered by Futamase and Maeda 
is theoretically consistent only if $\xi=1/6 >>10^{-3}$ and hence fine--tuning
in the initial conditions for the scalar field cannot be avoided. For
the particular value $\xi=1/6$, Futamase and Maeda gave an additional proof
that chaotic inflation cannot be realized \cite{FutamaseMaeda89}. 
The non--existence of inflationary solutions for the potential
(\ref{quarticpotential}) with a conformally coupled scalar field
and Einstein gravity was also pointed out in Ref.~\cite{AcciolyPimentel90}.

Chaotic inflation with the potential (\ref{quarticpotential}) for a
non--minimally coupled scalar field was considered   
in Ref.~\cite{MakinoSasaki91}. The purpose of that paper was to reduce the
fine--tuning of the parameter $\lambda$ in the potential imposed by
observations of the cosmic microwave background: $\lambda \leq 10^{-12}$. A
non--minimal coupling of the inflaton achieves this goal, but the price to be
paid is a fine--tuning in the value of the coupling constant: it has to be 
$| \xi | \simeq 10^4$ \cite{MakinoSasaki91}. However, the prescription 
$\xi=1/6$ to be applied to the model rules out this possibility.

Chaotic inflation with the potential
\begin{equation}
V( \phi)= \mu^2 \left( \frac{\phi^2}{2}+\frac{\lambda}{2n}\, \phi^{2n} 
\right)
\end{equation}
($\mu^2, \lambda >0$) and $\xi \neq 0$ was studied in Ref.~\cite{ALO90} with a
dynamical systems approach. Consistently with Ref.~\cite{FutamaseMaeda89},
the authors found that, for $\xi=1/6$, no trajectory in the phase 
space exists which corresponds to inflation \cite{ALO90}.

Chaotic inflation with the Ginzburg--Landau potential
\begin{equation}   \label{GinzburgLandau}
V( \phi)= \frac{\lambda}{8} \left( \phi^2-v^2 \right)^2
\end{equation}
and the Einstein Lagrangian for the pure gravity part of the action 
was considered in 
Ref.~\cite{FakirUnruh90}. In the case $\xi=1/6$, 
the authors of Ref.~\cite{FakirUnruh90} deduced that there are no inflationary
solutions. However, their analysis was performed in the regime $ \phi^2 
>>v^2$, in which the
potential reduces to the case of the quartic self--interaction 
(\ref{quarticpotential}). 

Chaotic inflation can be achieved 
in the context of induced gravity \cite{CervantesDehnen95}. No prescription 
for $\xi$ can be given in this case.

\subsection{Natural inflation}

In the natural inflationary scenario \cite{FreeseFriemanOlinto90}, the 
inflaton
is a massless pseudo Nambu--Goldstone boson with the potential
\begin{equation}
V( \phi )= \Lambda^4 \left[ 1+\cos \left( \frac{\phi}{f} \right) \right] \; ,
\end{equation}
which exhibits two energy scales: $f \sim m_{pl}$ and 
$\Lambda \sim 10^{-5}f$ is the scale of spontaneous symmetry breaking. This
scenario is motivated by superstring theories \cite{Turnerreview}, and
therefore there seems to be little indication on what prescription for $\xi$ is
correct, apart from the fact that GR is used in this scenario. This would imply
that inflation occurs in the low energy limit, in which the precription
$\xi=1/6$ applies. Since the analysis of the
potential is difficult, two regimes are considered \cite{Turnerreview}:
$i$)~$f \leq m_{pl}$, $V \simeq 2\Lambda^4$, which is extremely fine--tuned
\cite{SteinhardtTurner84}; $ii$) $f>>m_{pl}$, $V( \psi)=m^2 \psi^2 /2$ (where
$\psi=\phi-\sigma$, $\sigma=$constant), which is equivalent to the chaotic
inflationary scenario already considered.

\subsection{Double field inflation}

Inflation with two (or more) scalar field has been considered
\cite{Linde90b,Linde91PLB,AdamsFreese91,GarciaLinde95}; in 
Ref.~\cite{Linde91PLB} the potential is 
\begin{equation}
V( \phi,\psi)=\frac{\lambda}{4} \left( \psi^2-M^2 \right)^2+\frac{m^2}{2}
\, \phi^2+\frac{\lambda'}{2} \, \phi^2 \psi^2 \; .
\end{equation}
A particular realization for $\psi$ is a Peccei--Quinn field. Like in many
other papers, the theory considered is {\em classical}, but it is 
supposed to simulate the full quantum theory of the inflaton(s) by choosing 
a suitable potential. Such a hybrid theory may not be inconsistent for some
values of  the
coupling constant(s) of the scalar field(s) with the Ricci curvature. For
example, in Ref.~\cite{GarciaLinde95}, Brans--Dicke theory is used, and 
the inflaton is an extra (other than the Brans--Dicke) scalar field
non--minimally coupled to the curvature. The coupling constant is $\xi<0$,
$|\xi|<<1$. This scenario is inconsistent: in fact Brans--Dicke theory is a 
metric theory of gravity and any scalar field other than the Brans--Dicke field
is non--gravitational: therefore the EEP and the prescription 
$\xi=1/6$ apply to the inflaton in the scenario of Ref.~\cite{GarciaLinde95}.

In the scenario of Ref.~\cite{Copelandetal94} the Lagrangian for Einstein
gravity and two minimally coupled scalar fields are used. The theory 
is supposed to be ``a toy model for the scalar field
sector of the string--derived supergravity theory'' \cite{Copelandetal94}; in
supergravity there is no point in imposing the EEP (in fact the weak
equivalence principle is already violated in at least some 
realizations of the theory
\cite{Scherk}), which would guarantee the conformal coupling. We can only say
that, in GR, the minimal coupling for the two scalar fields of 
Ref.~\cite{Copelandetal94} is not acceptable, since they must be 
conformally coupled according to Ref.~\cite{SFFS}. The same 
conclusion applies to
the soft inflation of Ref.~\cite{BerkinMaedaYokoyama90}.

\subsection{Anisotropic cosmologies}

The occurrence of inflation has been studied also in anisotropic spaces for
a non--minimally coupled inflaton field. Starobinski \cite{Starobinski81} 
showed that, for $\xi=1/6$ (and therefore in GR), the anisotropic shear 
diverges as the inflaton
$\phi$ approaches the critical value $\phi_c=\left( 3/4\pi 
\right)^{1/2}m_{pl}$.
This result was recovered in Ref.~\cite{FutRotMat89}, in which it was also 
shown
that the divergence of the anisotropic shear also occurs if $\xi>0$ 
and for almost all initial conditions 
$\phi_0 >\phi_c$ (which do not reproduce the present universe). 
In general, the addition of anisotropy rules out the possibility of 
chaotic inflation for $ \xi >10^{-2}$ 
\cite{FutRotMat89}.
%%%%%%%%%%%%%%%%%%%%%%%%%%%%%%%%%%%%%%%%%%%%%%%%%%%%%%%%%%%%%%%%%%%%%%%%%%%
\section{Power--law inflation with the potential $V( \phi)=\lambda \phi^n$}

In Ref.~\cite{FutamaseMaeda89}, the case of a potential 
\setcounter{equation}{0}
\begin{equation} 
V( \phi)=\lambda \phi^n \;\;\;\;\; , \;\;\;\;\; n>6 
\end{equation}               
and $\xi \neq 0$ was considered: power--law inflation (\ref{PLI}) was
obtained, with
\begin{equation}      \label{qsta}
p=2\, \frac{ 1+(n-10) \xi}{(n-4)(n-6) | \xi |} \; .
\end{equation}
By substituting Eq.~(\ref{PLI}) in Eq.~(\ref{KG}), one obtains
\begin{equation}
\ddot{\phi}+\frac{3p}{t}\dot{\phi}+\frac{dV}{d\phi}+\frac{6\xi p (2p-1)}{t^2}
\, \phi=0   \; ,
\end{equation}
which has the solution 
\begin{equation}   \label{phiPLI}
\phi =\bar{\phi} \, t^{\alpha}
\end{equation}
where
\begin{equation}   \label{alpha}
\alpha=\frac{2}{2-n} \; 
\end{equation}
($\alpha<0$). It is to be noted that, in this particular case, it is possible
to write
\begin{equation}          \label{noidea}
\frac{\xi}{2} R \phi^2=\beta V( \phi) \; ,
\end{equation}
where
\begin{equation}
\beta = 3\xi p (2p-1) \lambda^{-1} {\bar{\phi}}^{2-n} \; .
\end{equation}
Usually, power--law inflation is associated to an exponential potential for a
minimally coupled field.
The possibility of obtaining power--law inflation with a
power--law potential is due to the non--minimal coupling of the inflaton to 
the Ricci curvature, and shows the effect of non--minimal coupling 
on the physics
of the scalar field and the dynamics of the universe. 
From Eq.~(\ref{noidea}), it may appear that one could 
substitute the physical system under consideration
with an ``equivalent'' Friedmann universe dominated by a 
minimally coupled scalar field with 
the effective potential
\begin{equation} \label{effectivepotential}
V_{eff}( \phi)=V( \phi)+\frac{\xi}{2}R\phi^2=\Lambda \phi^n \; ,
\end{equation}
where $ \Lambda=1+\beta$. However, Eqs.~(\ref{PLI}), (\ref{qsta}),
(\ref{phiPLI}) and (\ref{alpha}) do not constitute a solution of the coupled
Friedmann--Klein--Gordon equations for $\xi=0$. To see this, it is sufficient 
to consider the Friedmann equations for the ``equivalent universe''
\begin{equation}
\frac{\dot{a}^2}{a^2}=\frac{8\pi }{3} \rho \; ,
\end{equation}
\begin{equation}
\frac{\ddot{a}}{a}=-\frac{4\pi}{3} \left( \rho +3P \right) \; ;
\end{equation}
these, together with Eqs.~(\ref{density}), (\ref{pressure}) for $\xi=0$ and 
(\ref{alpha}), (\ref{effectivepotential}) give $p
\propto t^{4/ (2-n)}$, which contraddicts the constancy of $p$. Therefore, 
the
introduction of the effective potential (\ref{effectivepotential}) is not
useful even when Eq.~(\ref{noidea}) is satisfied. The conformal technique 
used in Ref.~\cite{Kaiser} appears more promising.

In order to achieve inflation, it must be $p>1$.  In
the space of parameters $(n,\xi)$, the inequality $p>1$ is satisfied 
only in the regions:
\begin{equation}
n>6 \;\;\;\; , \;\;\;\; 0<\xi <\frac{2}{n^2-12n+44} \; ,
\end{equation}
\begin{equation}
6<n<4+2\sqrt{3}\simeq 7.464 \;\;\;\; , \;\;\;\;\xi<0 \; ,
\end{equation}
\begin{equation}
n=4+2\sqrt{3} \;\;\;\; , \;\;\;\;\xi < \frac{1}{4(3-\sqrt{3})}\simeq 0.197
  \; ,                                      \end{equation}
\begin{equation}
n> 4+2\sqrt{3}  \;\;\;\; , \;\;\;\;\frac{-2}{n^2-8n+4}<\xi<0 \; .
\end{equation}
The range of values $6\leq n \leq 10$ is interesting 
for superstring theories \cite{Shafi93}; only a very narrow range of values of
$\xi$ is allowed for high $n$. However, it must be kept in mind that 
fine--tuning arguments 
rule out the scenario for $\xi >0$ \cite{FutamaseMaeda89}. 
%%%%%%%%%%%%%%%%%%%%%%%%%%%%%%%%%%%%%%%%%%%%%%%%%%%%%%%%%%%%%%%%%%%%%%%
\section{Observational constraints on the coupling parameter $\xi$}

As discussed in the previous section, many inflationary scenarios are not
viable for certain ranges of values of $\xi$. Other scenarios are viable, with
the parameter $\xi$ spanning a range of values, which can be constrained by
the available observations of cosmic microwave background anisotropies.

Kaiser \cite{Kaiser} considered chaotic inflation with the potential $V(
\phi)=\lambda \phi^4$ and a non--minimally coupled scalar field, and computed
the spectral index of density perturbations  as a function 
of $\xi$ \cite{footnote7}:\setcounter{equation}{0}
\begin{equation}        \label{nkaiser1}
n_s=1-\frac{32 \xi}{1+16\alpha \xi} \; ,
\end{equation}
where $\alpha$ is the number of e--folds of the scale factor before the end
of inflation. In the following, we will use the value $\alpha=60$ adopted in
Ref.~\cite{Kaiser}. This model is different from the one given by 
Eqs.~(\ref{PLI}) 
and (\ref{qsta}). The statistical analysis of data from the {\em COBE} 
experiment detecting anisotropies in the cosmic microwave background gives
$n_s=1.1\pm 0.5$ \cite{Smootetal}, and the combined statistical analysis 
of the
{\em COBE} and Tenerife observations yields the 1$\sigma$ limit $n \geq 0.9$
\cite{Hancocketal94}. We adopt the limits
\begin{equation}     \label{nlimits}
0.9 \leq n_s \leq 1.6
\end{equation}
which, using Eq.~(\ref{nkaiser1}) yields the constraints on $\xi$:
\begin{equation}
\xi \leq -1.56 \cdot 10^{-3} \;\;\;\;\;\;\; , \;\;\;\;\;\;\; \xi \geq -9.87
\cdot 10^{-4} \; .
\end{equation}
The GR prediction $\xi=1/6$ implies $n_s=0.967$. However, values of $\xi$
greater than $\sim 10^{-3}$ lead to fine--tuning problems
\cite{FutamaseMaeda89,AcciolyPimentel90,MakinoSasaki91,ALO90}, as explained in
Sec.~3.

Chaotic inflation with a non--minimally coupled scalar field and the
Ginzburg--Landau potential (\ref{GinzburgLandau}) was also considered in
Ref.~\cite{Kaiser}. The spectral index of density perturbations was 
computed in the
two regimes: $a)$ $\phi_{end}^2>>v^2$; $b)$ $ \phi_{end}^2 \simeq v^2$,
respectively, where $\phi_{end}$ is the value of the scalar field at the end
of inflation. Case $a)$ is reduced to the case, already considered, of a
quartic potential and of Eq.~(\ref{nkaiser1}). Case $b)$ yields \cite{Kaiser}
\begin{equation} \label{nkaiser2}
n_s( \xi,\delta )=1-\frac{16 \xi (1+\delta^2 )}{8\alpha \xi 
(1+\delta^2)-\delta^2} \; .
\end{equation}
From Eqs.~(70) and (71) of Ref.~\cite{Kaiser}, one derives
\begin{equation}  \label{delta}
\delta^2 ( \xi, v)=-\, \frac{8\pi G \xi v^2}{1+8\pi G \xi v^2} \; ,
\end{equation}
where $G=m_{pl}^{-2}$ is the present value of Newton's constant. Although
$n_s$ was given in Ref.~\cite{Kaiser} for a range of values of 
$\delta$ and $\xi$,
it turns out that $n_s$ depends only on the square of the parameter $v$ 
and not from $\xi$ \cite{footnotereferee}.
Using Eq.~(\ref{delta}), one obtains
\begin{equation}  \label{nn}
n_s (v)=1-\frac{2}{\alpha+\pi \left( v/ m_{pl}\right)^2} \; .
\end{equation}
The limits (\ref{nlimits}) are satisfied for all values of $v$, hence 
Eq.~(\ref{nn}) does not constrain the parameter $v$.

The last scenario considered in Ref.~\cite{Kaiser} is the case of 
a non--minimally
coupled scalar field, the Ginzburg--Landau potential (\ref{GinzburgLandau}),
$\phi_{end}^2  \simeq v^2$ and new--inflationary initial 
conditions, which give
\begin{equation}
n_s=1+8\xi \,  \frac{1+\delta^2}{\delta^2} \; .
\end{equation}
Again, the use of Eq.~(\ref{delta}) reveals that $n_s$ is independent of $\xi$
and is a function of $v^2$ only:
\begin{equation}   \label{nnn}
n_s(v)=1-\frac{1}{\pi \left( v/m_{pl} \right)^2} \; .
\end{equation}
The limits (\ref{nlimits}) provide a constraint on the parameter $v$ of the
Ginzburg--Landau potential:
\begin{equation}
|v| \geq \left( \frac{10}{\pi} \right)^{1/2} m_{pl} \simeq 1.78 \,m_{pl} \; .
\end{equation}
We are not aware of other viable scenarios in which the spectral index $n_s$
has been computed as a function of the coupling constant $\xi$. In 
the last two
scenarios considered in this section, the independence of $n_s$ of $\xi$
rules out the possibility of determining this parameter with 
the data currently
available.

\section{Discussion and conclusions}

The problem of the correct value of the coupling parameter $\xi$ in any given
inflationary scenario containing scalar fields cannot be neglected, if the
scenario is to be theoretically consistent. We have analyzed the inflationary
scenarios which are most studied in the literature: some of them are
theoretically consistent, while some others are not, and in other cases the
value (or range of values) of $\xi$ is unknown. A clear prescription for the
value of $\xi$ emerges in GR, and it has been shown that GR is an
attractor for scalar--tensor theories 
\cite{DamourNordvedt,GarciaQuiros90,GBW}. If
scalar--tensor theories approach GR during the matter--dominated epoch of the
universe, as suggested in Ref.~\cite{DamourNordvedt}, these 
arguments are irrelevant
for inflationary scenarios. It has also been proposed  that the ``GR as an
attractor'' behavior occurs {\em during} inflation \cite{GarciaQuiros90,GBW}; 
in this case
the coupling parameter $\xi$ assumes the value $1/6$ before the end of
inflation. The relevance of this phenomenon depends on the time during
inflation at which the scalar--tensor theory approaches GR, and is worth 
studying in the future. 

It is also to be remarked that, if GR is the correct theory of gravity during
inflation, or if the inflaton field is conformally coupled to the Ricci
curvature in some other theory
of gravity, the universe has a peculiar feature: the cosmological tail problem
\cite{tails} for the $\phi$--field (i.e. the backscattering off the
background curvature of spacetime) is trivially resolved in some cases: due 
to the conformal flatness of the Friedmann universe and to the conformal
invariance of the Klein--Gordon equation, a massless scalar field with 
potential $V=0$ or $V=\lambda \phi^4$ (chaotic inflation) 
propagates withouth tails.

It should also be kept in mind that, in the inflationary scenarios not based on
GR, there is, in principle, the possibility (not explored so far) that the
inflaton couples non--minimally to scalars constructed from the Riemann tensor
and different from $R$. These couplings are not allowed in GR, or in any metric
theory of gravity in which a non--gravitational inflaton satisfies
Eq.~(\ref{KG}).

Finally, we remark that it is believed that particles associated to the
inflaton field may survive as dark matter in boson stars. In this case, the
correct value of the coupling constant $\xi$ in a specific inflationary
scenario must also be used in the study of the structure and stability of 
boson stars, both of which depend on $\xi$. Another possible 
application of the
prescriptions of Sec.~2 is the field of classical and quantum wormholes.

\section*{Acknowledgments}

The author thanks M. Bruni, S. Sonego and F.I. Cooperstock 
for stimulating discussions. An anonymous referee is also acknowledged for
suggestions leading to improvements in the manuscript. 
This work was supported by a NATO Advanced Fellowship Programme 
through the National Research Council of Italy (CNR). 


\begin{thebibliography}{99}
\bibitem{footnote1} Other notations and conventions used in this paper are as
follows: the metric signature is --~+~+~+. The speed of light $c$ and the
(reduced) Planck constant assume the value unity. 
$m_{pl}=G^{-1/2}$ is the Planck mass and $G$ is Newton's constant.
Greek indices assume the
values 0,~1,~2,~3, Latin indices assume the values 1,~2,~3. The Ricci tensor
is given in terms of the Christoffel symbols $\Gamma_{\alpha\beta}^{\delta}$ 
by $R_{\mu\rho}=
\Gamma^{\nu}_{\mu\rho ,\nu}-\Gamma^{\nu}_{\nu\rho ,\mu}+
\Gamma^{\alpha}_{\mu\rho}\Gamma^{\nu}_{\alpha\nu}-
\Gamma^{\alpha}_{\nu\rho}\Gamma^{\nu}_{\alpha\mu} $, and 
$\Box \equiv g^{\mu\nu}\nabla_{\mu}\nabla_{\nu}$.
\bibitem{footnote2} In the early works on 
inflationary cosmology, a Higgs field was
believed to be responsible for inflation. This picture has changed, but a Higgs
field is still a candidate to the role of the inflaton in some scenarios. For
example, the simplest model of hybrid inflation \cite{Linde91PLB} identifies
one of the two scalar fields in it as the Higgs field and correctly uses
$\xi=0$.
\bibitem{footnote4} Despite this fact, the experimental tests of Brans--Dicke 
theory still refer to the Jordan frame formulation \cite{Will}.
\bibitem{footnote5} Although the effect is like that of a mass, 
the interpretation of the constant curvature
as a mass term for the scalar field must not be taken literally
\cite{FaraoniCooperstock95}.
\bibitem{footnote6} These particular models,
however, are strongly constrained by the available observations of 
cosmic microwave
background anisotropies \cite{FaraoniCooperstockOverduin}.
\bibitem{footnote7} The convention on the sign of $\xi$ in Ref.~\cite{Kaiser} 
is the opposite of ours.
\end{thebibliography}
\end{document}